\documentclass{epsconf}
\usepackage{graphicx}
\usepackage{wrapfig}
\usepackage{amsmath}
\usepackage[english]{babel}
\usepackage{color}

\newcommand{\ie}{\emph{i.e., }}

\newcommand{\figref}[1]{Fig.\ref{#1}}

\title{Energetic particle transport:\\ diffusion vs convection and phase-space barriers}
\author{\underline{N. Carlevaro}$^{1}$, G. Montani$^{1,2}$, M.V. Falessi$^{1}$, Ph. Lauber$^{3}$}
\institute{$^{1}$ ENEA, FNS Department, C.R. Frascati,
Via E. Fermi 45, 00044 Frascati (Roma), Italy\\
$^2$ Physics Department, ``Sapienza'' University of Rome, P.le Aldo Moro 5, 00185 Roma, Italy\\
$^3$Max-Planck-Institut fuer Plasmaphysik, Boltzmannstrasse 2, D-85748 
Garching, Germany
}

\begin{document}
\maketitle

\vspace{-0.8cm}
\paragraph{Abstract}
Energetic particle (EP) redistribution \cite{cz,bs} in the presence of multiple Alfv\'en eigenmodes (AEs) is analyzed in Ref.\cite{spb16} for the ITER 15MA baseline scenario: non-linear hybrid simulations (within their well known limits) point out
that transport can be dominated by avalanches under certain conditions. These phenomena are properly reproduced by the 1D reduced description of Ref.\cite{ppcf}. Here, using this simplified 1D model, we define the transport character (convective/diffusive) of self-consistent EP redistribution. Transport barriers in phase space are studied using the Lagrangian Coherent Structures (LCS) technique \cite{mvf}.

\vspace{-5mm}
\paragraph{Reduced model for EP transport} In Ref.\cite{ppcf}, a mapping procedure has been set up to translate the evolution of the radial profile of fast ions (interacting with multiple AEs) into the dynamics of an equivalent beam-plasma system (BPS). The mapping consists in a one-to-one link between the radial coordinate ($s$) of the realistic 3D scenario and the 1D BPS velocity ($u$). It is defined around a single chosen resonance, resulting in a linear relation, and then extended to the multiple mode case. For the reference resonance, the mapping is formally constructed locally from the resonance condition and reads as: $u=(1-s)/\ell^*$ (where $\ell^*$ is an arbitrary constant defining the spectral features and the periodicity length of the 1D slab).

The BPS setup can be closed by determining the beam to plasma density ratio parameter ($\eta$). We note that the EP/AE system has an higher dimensionality with respect to the BPS and, thus, the non-linear BPS transport results more efficient than the realistic radial transport, in the case of equal drive \cite{ppcf}. $\eta$ can be set to obtain optimized EP relaxation in the $s$ direction: 
we fix such parameter by performing a scan in order to match the relaxed QL flattening width of the two schemes directly integrating QL equations for the BPS \cite{ql20nc}. Multiple modes can be included by means of the resonance conditions written in the radial space and, to preserve the asymptotic mode decay, we impose equal dampings (normalized to the effective mode frequencies $\omega$ for the BPS and $\Omega$ for the 3D system) between the two systems. Mode frequencies and drives of the reduced scheme can be in fact evaluated using the linear BPS dispersion relation considering the initial averaged 3D EP distribution mapped in $u$. The dispersion relation also provides, as a result, the drives $\gamma_{Dj}$ of the reduced model. Due to the reduced dimensionality and to the effective wave-particle power exchange, $\gamma_{Dj}$ will result smaller than the normalized linear AE drives $\Gamma_{Dj}$. Introducing $\gamma_{Dj}/\omega_j=\alpha_j\;\Gamma_{Dj}/\Omega_j$, we can evaluate the important parameter $\alpha_j (\leq1)$ indicating, roughly, the fraction of the most resonant particles. In fact, $\alpha\simeq1$ when comparing only EPs which maximize the energy exchange in the realistic scenario. In Ref.\cite{ppcf}, it also traced the methodology for estimating the parameter $\alpha$ by invoking the phase space splitting.

\begin{figure}[ht!]
\centering
\includegraphics[width=0.27\linewidth]{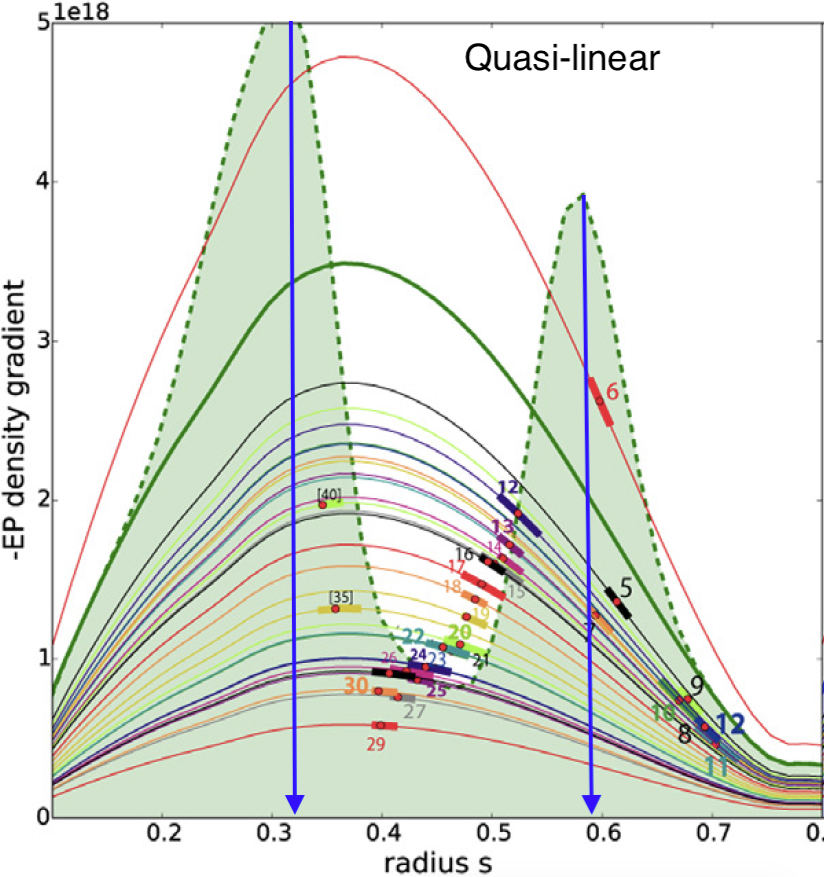}
\includegraphics[width=0.44\linewidth]{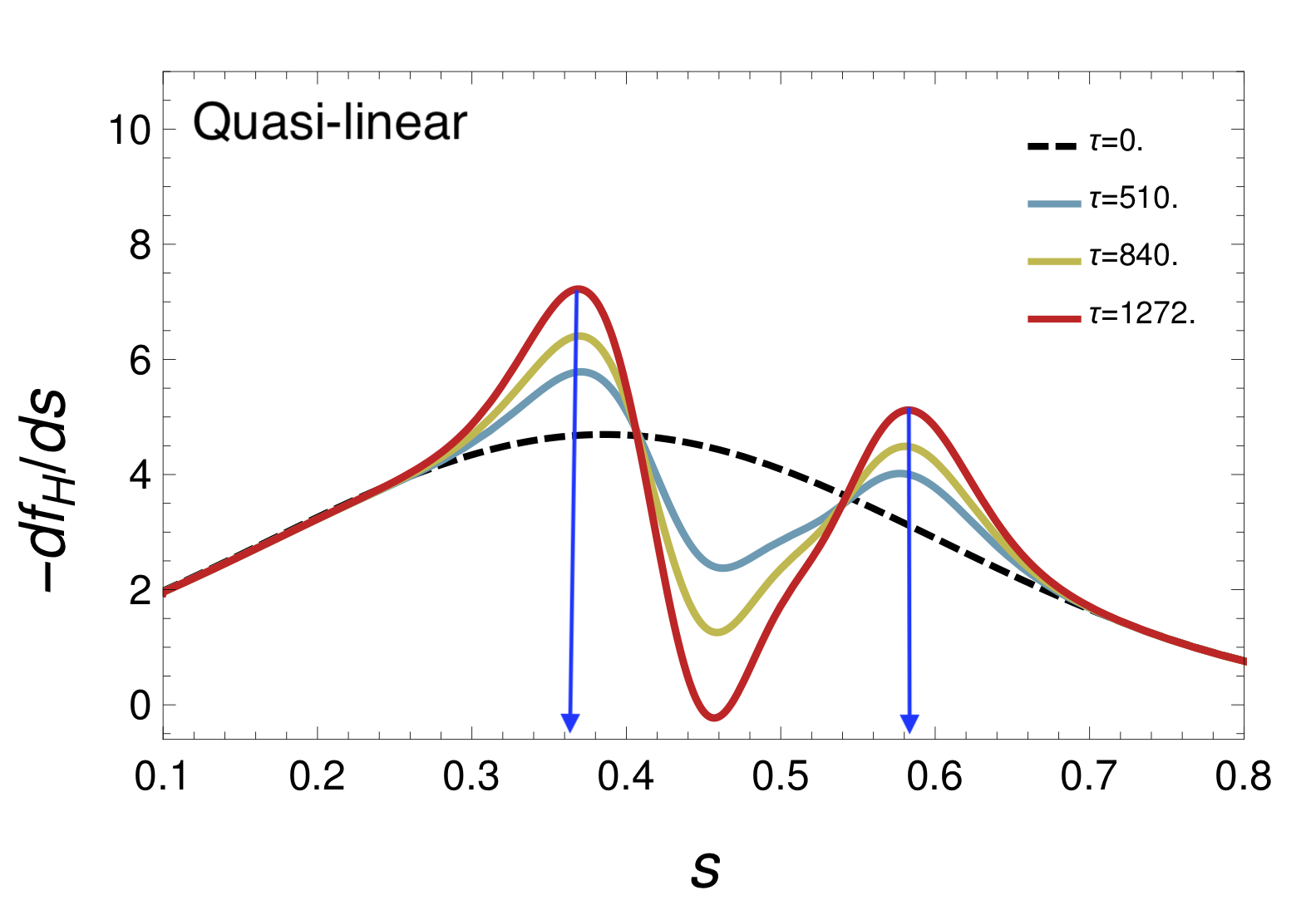}
\vspace{-4mm}
\caption{\it \footnotesize Gradient of the EP redistribution from QL theory. Left-hand panel: results of Ref.\cite{spb16} for late times. Right-hand panel: reduced BPS evolution. Blue arrows indicate the flattening width.
\label{fig1}}
\end{figure}

\begin{figure}[ht!]
\centering
\includegraphics[width=0.4\linewidth]{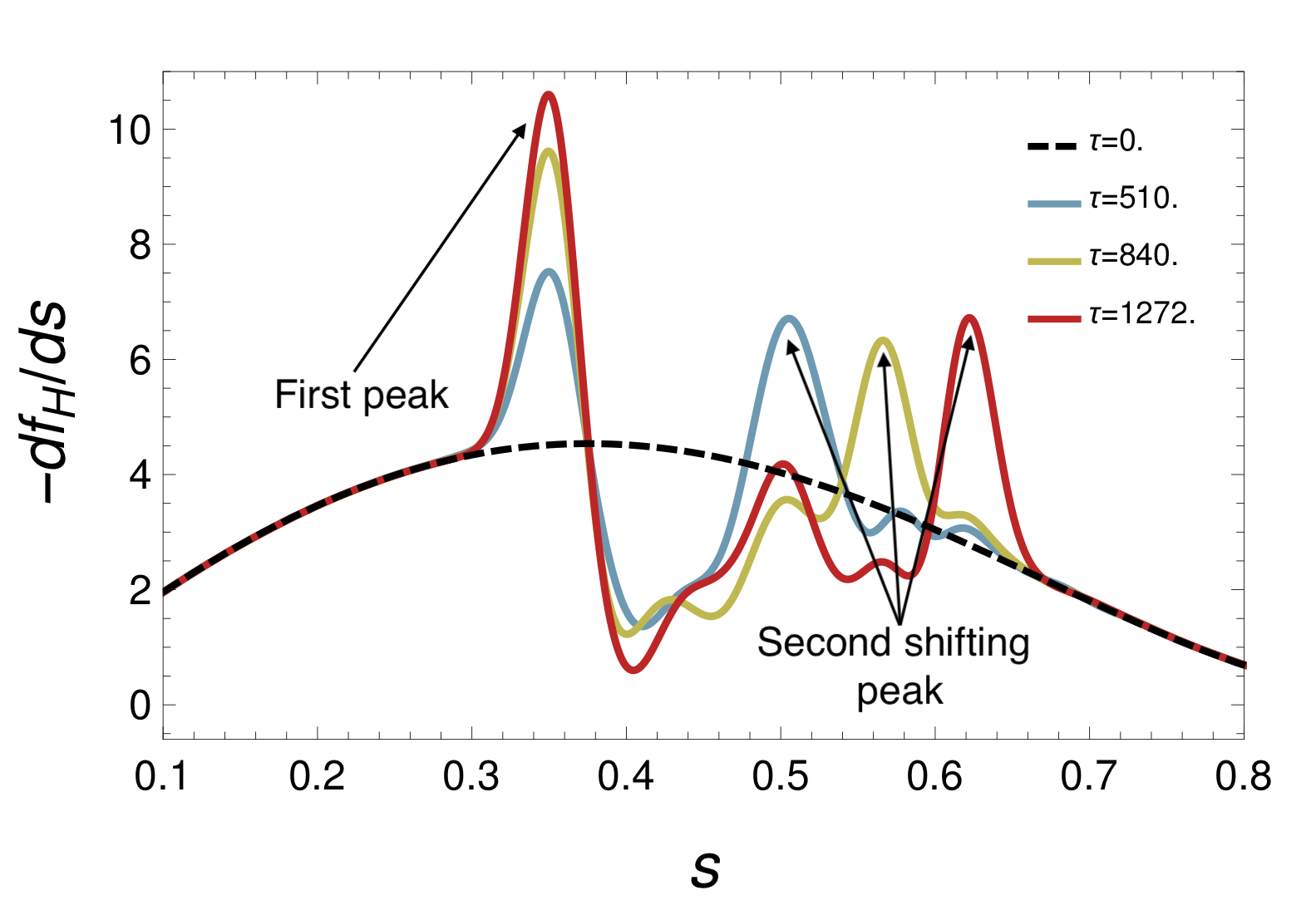}
\includegraphics[width=0.4\linewidth]{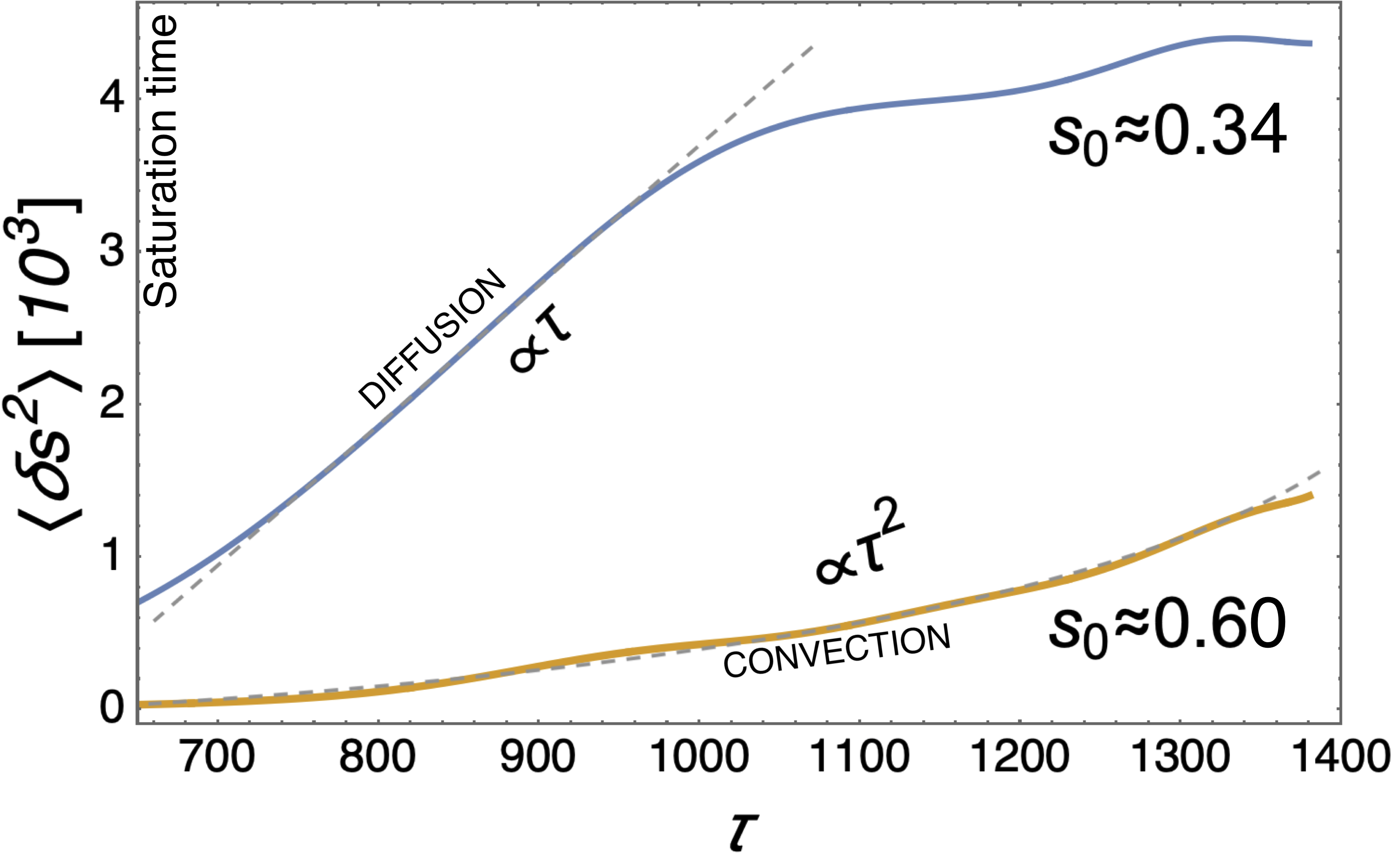}
\vspace{-7mm}
\caption{\it \footnotesize Left-hand panel: evolution of EP profile gradient as function of $s$. Colored lines represent different times (dashed black line is the initial distribution $f_{H0}$). Right-hand panel: plot of $\langle\delta s^2\rangle(\tau)$ for tracers initialized around $s_0\simeq0.34$ (blue) and $s_0=0.6$ (yellow). Dashed gray lines are guide for eyes underlining the two regimes.
\label{fig2}}
\end{figure}

\vspace{-5mm}
\paragraph{EP redistribution}
We analyze the ITER 15MA beseline scenario described in Ref.\cite{spb16} by runnig simulations of the BPS mapped back into the radial coordinate $s$. The addressed scenario includes the least damped 27 toroidal AE: $n\in[12,30]$ for the main branch and $n\in[5,12]$ for the low branch (the role of the poloidal harmonics is discussed in Ref.\cite{ppcf}). $n=21$ is chosen as reference resonance for the mapping since it is characterized by average value of radial position and essentially also of the growth rate of the main branch. QL evolution presented in Ref.\cite{spb16} outlines a flattening width of the distribution function $f_H$ between $s\simeq0.32$ and $s\simeq0.58$. After a scan by integrating the BPS version of the QL equations, this width is found to be reproduced for $\eta\simeq0.007$ (see Fig.\ref{fig1}) corresponding to $\alpha_{21}=0.4$  (we have set $\ell^*=400$).\vspace{-0.9cm}\\
\vspace{-0.0cm}\\
The reduced self consistent evolution of the distribution function is now plotted the left-hand panel of Fig.\ref{fig2}. While a first peak is well localized around $s\simeq0.35$ (excitation of the high branch), a second peak, related to the linearly stable low branch, is shifting in time toward $s\simeq0.65$, outlining an avalanche-like transport toward the edge. This sheds light on the relevance of the stable (sub dominant) spectrum component destabilization due to domino EP dynamics.

\vspace{-5mm}
\paragraph{Diffusion vs convection}
We now study the transport character during the self-consistent relaxation by means of test particle evolution. The tracers are initialized around a reference radial values $s_0$ to characterize different portions of the distribution function and the transport features are defined by the mean square path as a function of time (average $\langle...\rangle$ taken over the tracers):
\begin{align}\nonumber
\langle\delta s^2\rangle=\langle[s(\tau)-\langle s(\tau)\rangle]^2\rangle\;,\quad \langle\delta s^2\rangle\propto\tau \Rightarrow\text{DIFFUSION}, \quad
\langle\delta s^2\rangle\propto \tau^2 \Rightarrow \text{CONVECTION}.
\end{align}\vspace{-2.3cm}\\
\vspace{-0.4cm}\\
In \figref{fig2} (right-hand panel), we plot $\langle\delta s^2\rangle$ as a function of time for tracers initialized around $s_0=0.34$ (close to the first peak), and around $s_0=0.6$ (second peak). Focusing on time scales after saturation (\ie after the dynamics drive dominance by rotating phase space clumps), the diffusive behavior of tracers initialized in the first peak is evident during the post saturation phase. The late dynamics does not outline pure diffusive features due to the statistical weight of significantly displaced tracers. Tracers of the second peak are instead globally characterized by a $\tau^2$ dependence during the whole considered time interval, underlining the convective character of the transport due to the dominance of avalanches related to the low-$n$ branch.

\vspace{-5mm}
\paragraph{Phase-space transport barriers}
The local dynamics is related to phase space rotating clumps. These can be identified by studying the regions where the mixing phenomena are faster and their borders which are called Lagrangian Coherent Structures (LCS) \cite{mvf,ncjpp}: the most repulsive or attractive material line defined by a peaked profiles of the Finite Time Lyapunov Exponent (FTLE) fields. We adopt here a test particle approach by tracing trajectories of markers initialized, at a given time $\tau$, in two phase-space grids (the 1D BPS phase space $(\bar{x},u)$, where $\bar{x}$ denotes the normalized ``space'' coordinate, is mapped into $(\bar{x},s)$): one defined by $(\bar{x}_i,s_j)$ and a second radial displaced one at $(\bar{x}_i,s'_j)$, where $s'_j=s_j+\delta_{ij}$ with $\delta_{ij}=\delta=const$. At $\tau+\Delta\tau$, the markers will be at a distance $\Delta_{ij}$ and the FTLE field in the phase space $(\bar{x},s)$ is defined as 
\begin{equation}\label{ftleeq}
\sigma(\bar{x}_i,s_j,\tau)=
\ln\,[\Delta_{ij}/\delta]/\Delta\tau\;.
\end{equation}
For $\Delta\tau>0$ ($<0$), the FTLE peaked profiles define repulsive (attractive) transport barriers.
\begin{figure}[ht!]
\includegraphics[width=0.32\linewidth]{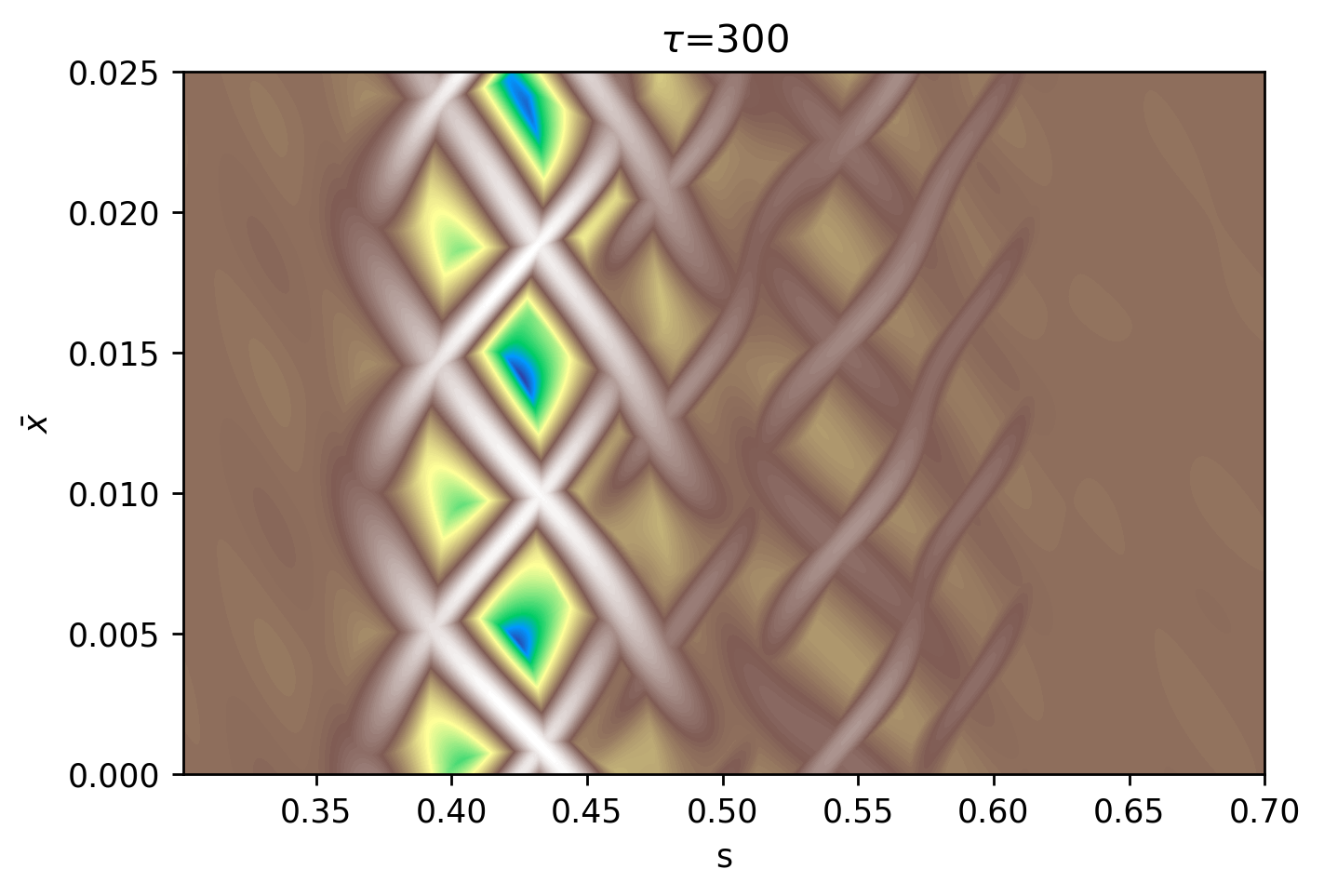}
\includegraphics[width=0.32\linewidth]{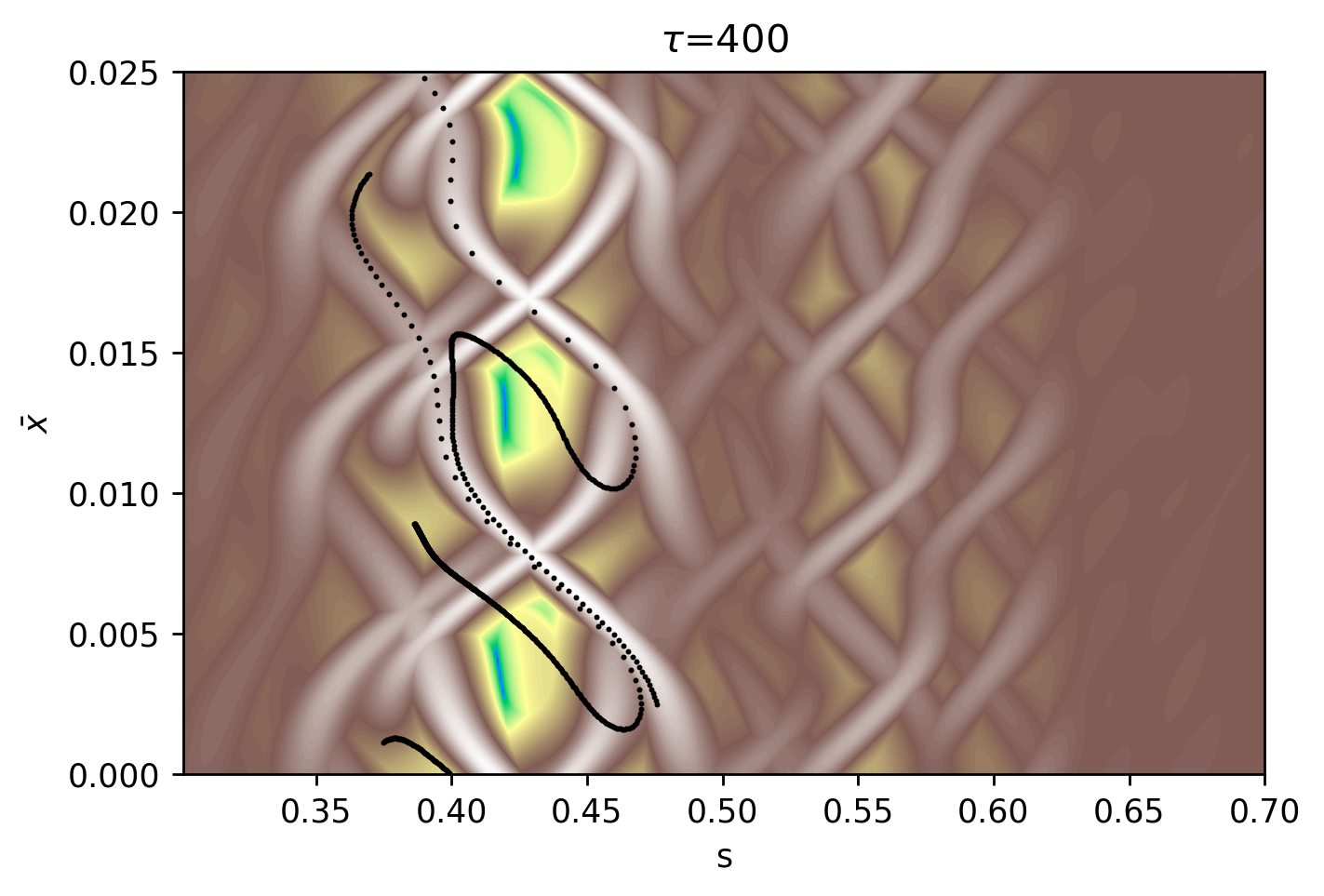}
\includegraphics[width=0.32\linewidth]{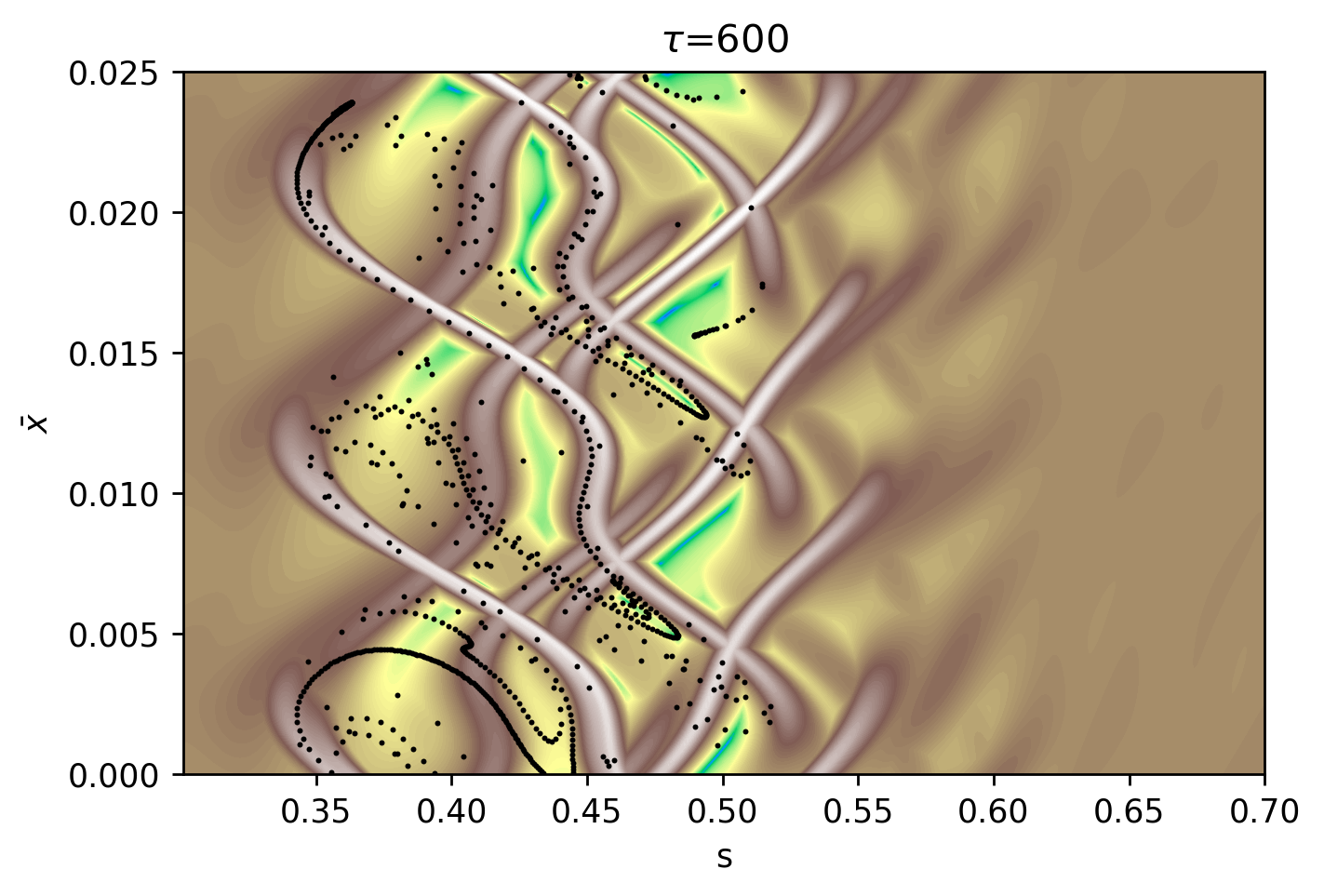}\\
\includegraphics[width=0.32\linewidth]{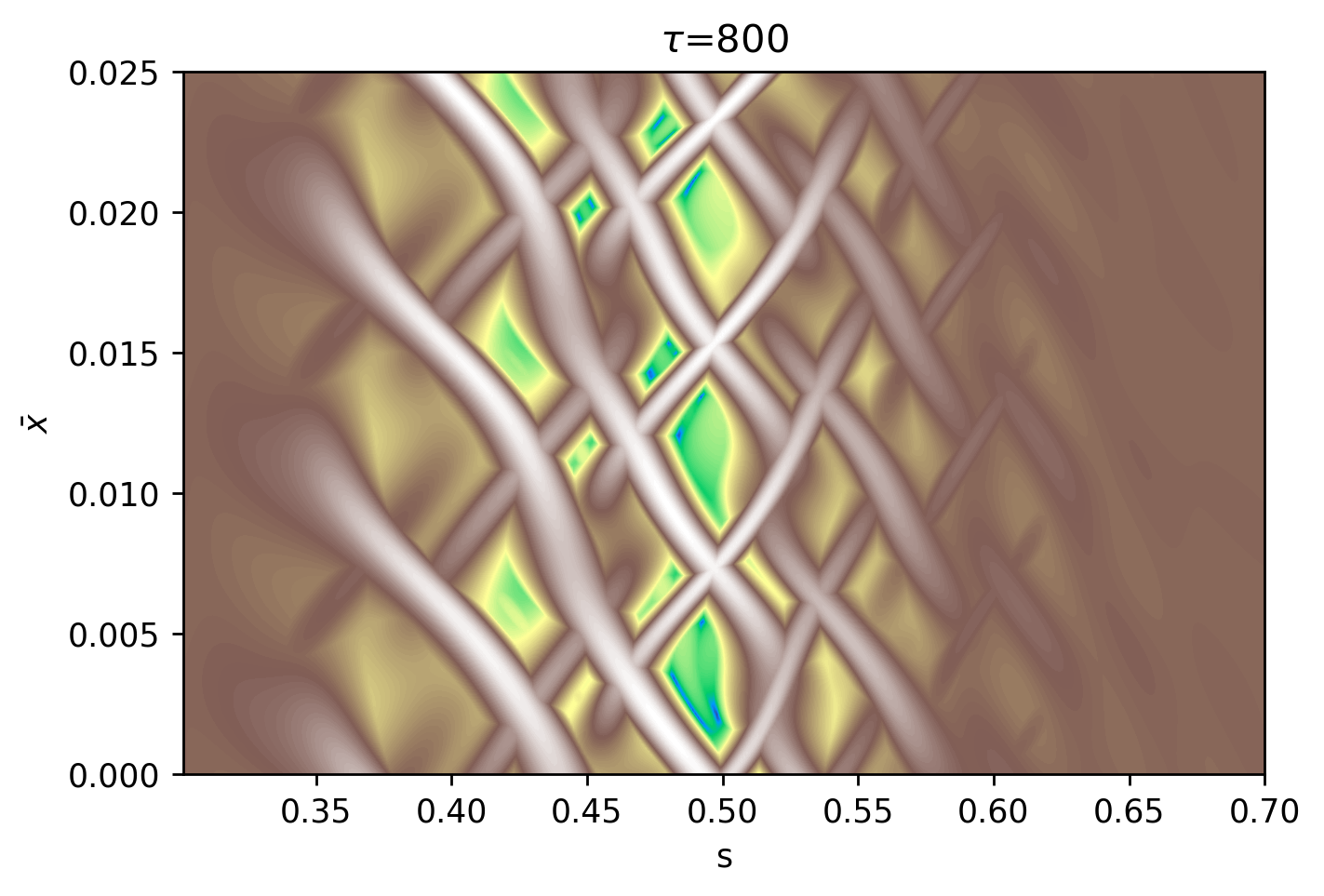}
\includegraphics[width=0.32\linewidth]{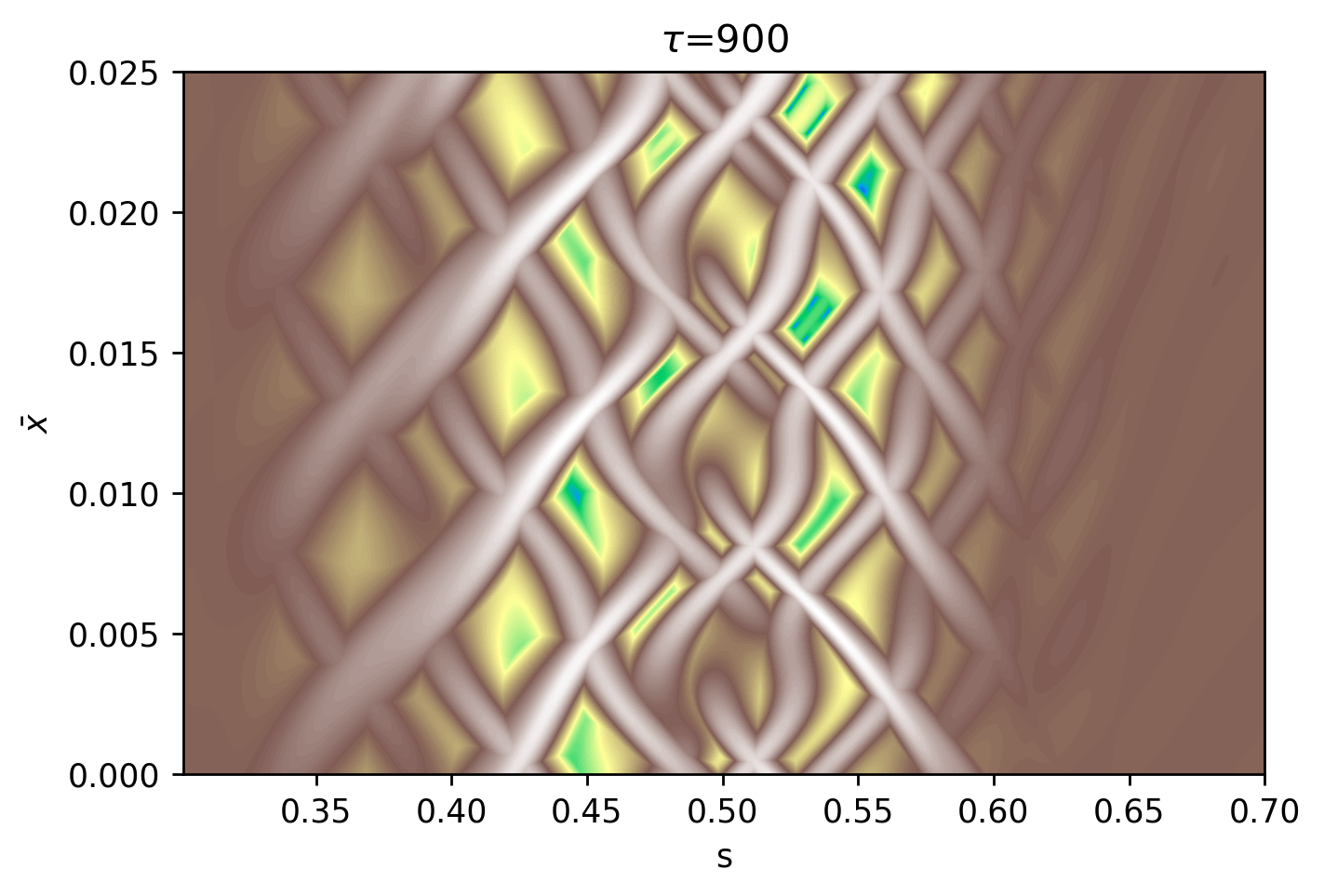}
\includegraphics[width=0.32\linewidth]{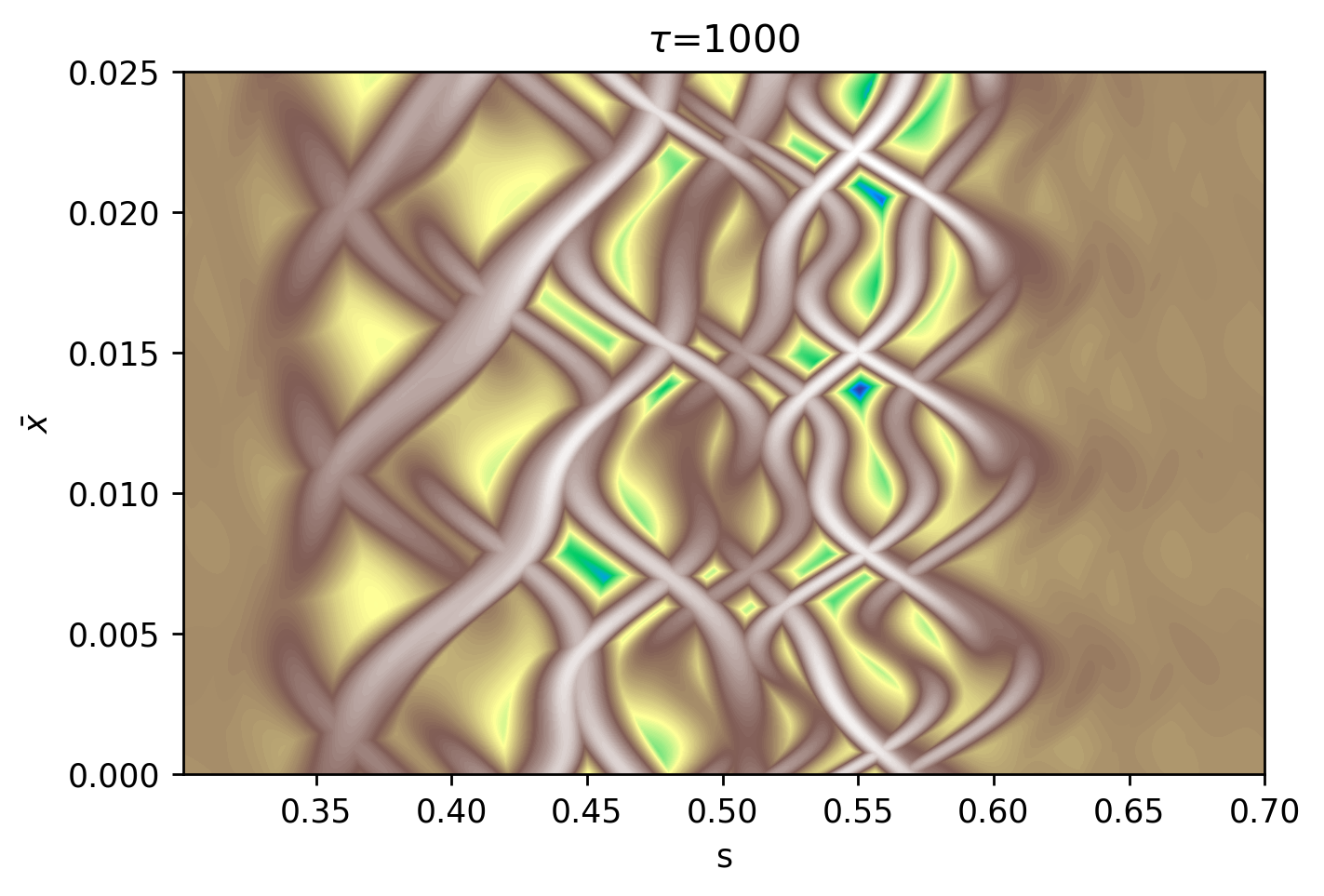}\\
\includegraphics[width=0.32\linewidth]{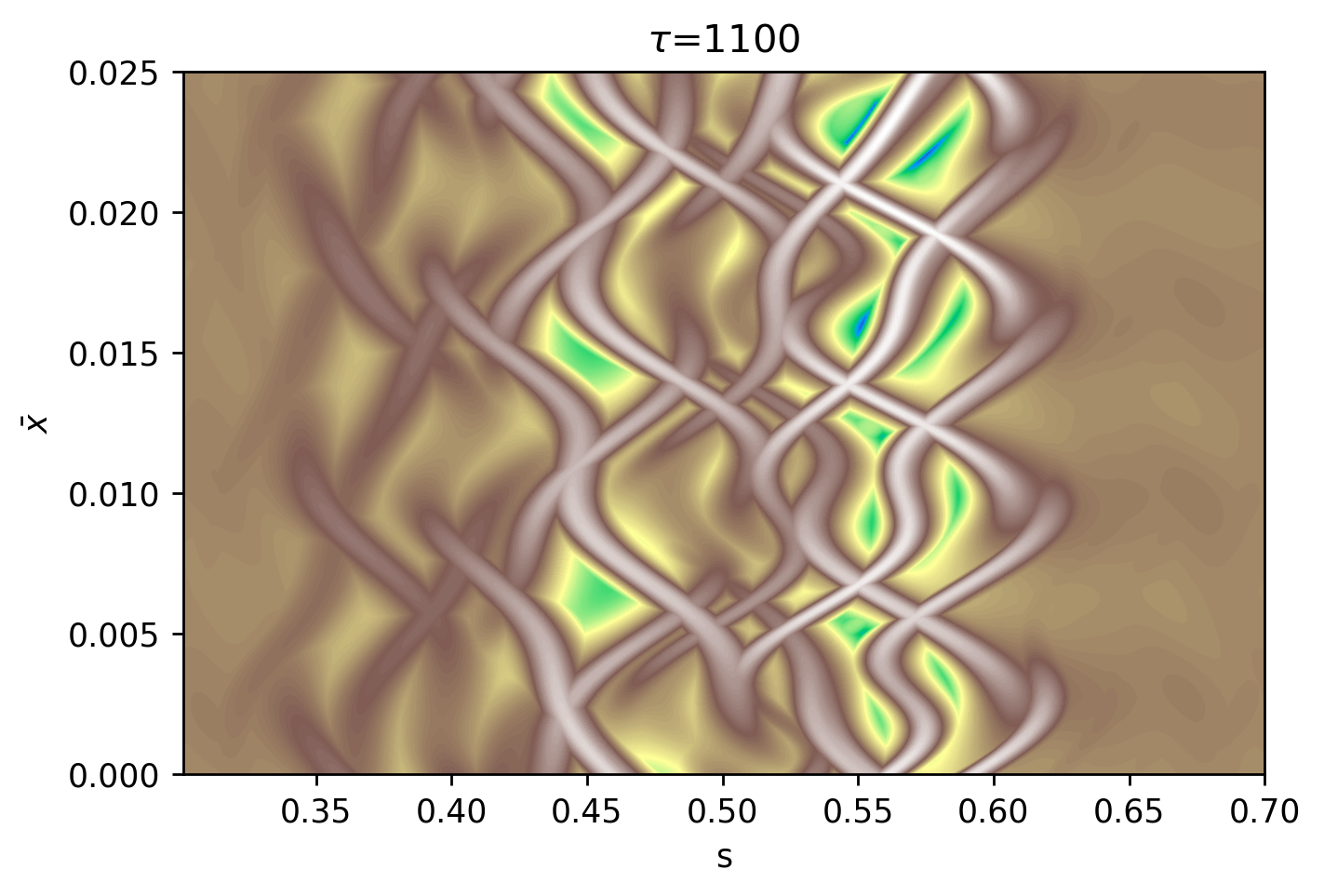}
\includegraphics[width=0.32\linewidth]{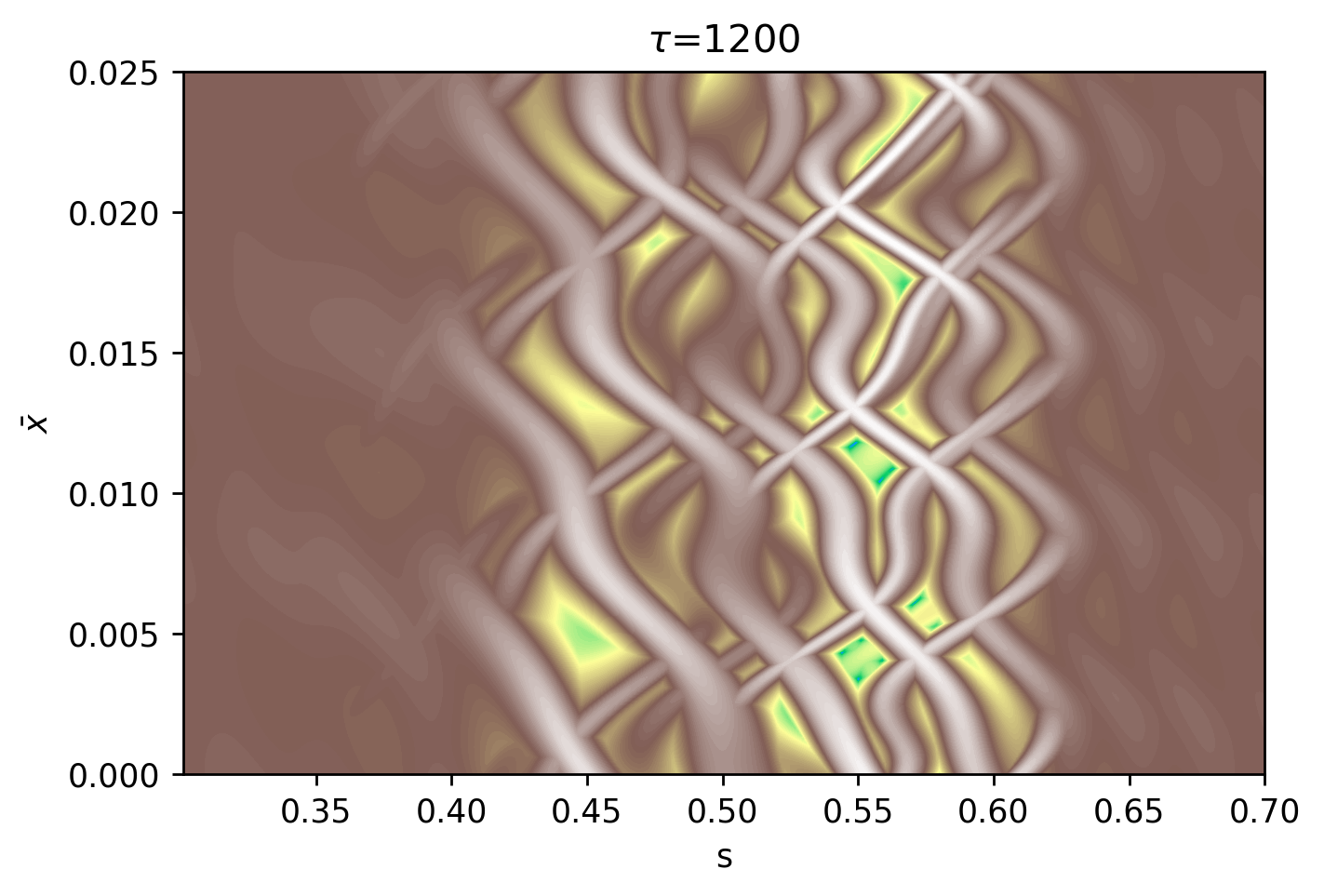}
\includegraphics[width=0.32\linewidth]{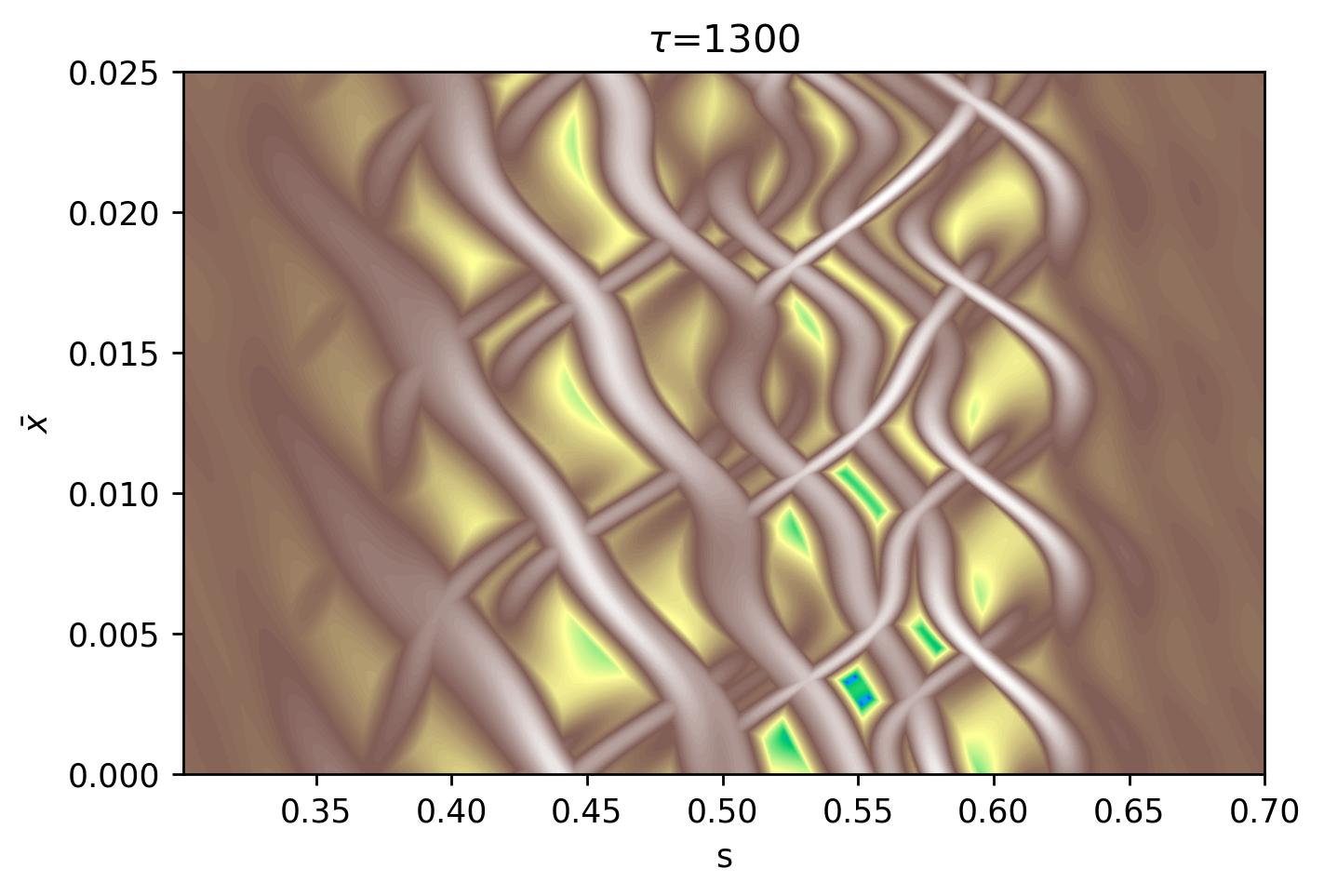}
\vspace{-5mm}
\caption{\it \footnotesize Contour plots of $\sigma(\bar{x}_i,s_j)$ from Eq.(\ref{ftleeq}) at fixed times as indicated over the graphs. We set $\Delta\tau=\pm100$ and we use arbitrary units with white/brown corresponding to peaks, while green/blue denote valleys. For the snapshots at $\tau=400,600$, we also overplot a sample of tracers (in black) to highlight attractive/repulsive barriers.   
\label{figftle}}
\end{figure}

In \figref{figftle}, we show the contour plots of $\sigma(\bar{x}_i,s_j)$ at fixed times implementing two grids, each with $1.6\times10^5$ tracers, with $\Delta\tau=\pm100$. For two snapshots, we overplot sample markers in order to underline the early rotating clump dynamics and also to show the attractive/repulsive character of the barriers. At early times, the dominance of the high branch modes ($n\in[12,30]$) is consistent with the more peaked transport barriers in the corresponding resonance region at low $s$.  As time increases, the progressive avalanche excitation of the low branch ($n\in[5,12]$) yields to the emergence of more robust barriers moving toward large $s$ values.

\vspace{-5mm}
\paragraph{Concluding remarks}
We showed how the BPS translation of a realistic
3D scenario for the interaction of EP with AE be predictive in reproducing fundamental transport features. In particular, we were able to characterize the self-consistent dynamics by determining diffusive and convective transport regimes. The analysis on the LCS allowed a characterization of the transport barrier formation and disruption in the phase space. This analysis offers a valuable background for future investigations concerning the mapping here considered, in view of a precise algorithm to estimate EP fluxes from very fast reduced numerical codes.

\tiny This work has been carried out within the framework of the EUROfusion Consortium [ER Project ATEP - CfP-FSD-AWP21-ENR-03-MPG-01], funded by the European Union via the Euratom Research and Training Programme (Grant Agreement No 101052200 — EUROfusion). Views and opinions expressed are however those of the author(s) only and do not necessarily reflect those of the European Union or the European Commission. Neither the European Union nor the European Commission can be held responsible for them.

\end{document}